\documentclass[prl, twocolumn, showpacs, amsmath, amssymb,superscriptaddress]{revtex4-1}
\usepackage{amssymb,amsfonts,amsmath} 
\usepackage{graphicx}
\usepackage{color}
\usepackage{hyperref}

\usepackage{longtable}
\usepackage{bbm}
\usepackage{ulem}
\usepackage{enumitem}
\usepackage{comment}
\usepackage{times}
\usepackage{nicefrac}

\definecolor{augustine}{rgb}{1,0.5,0.5}
   
\definecolor{dante}{rgb}{0,1,1}

\begin{document} 

\title{Stark time crystals: Symmetry breaking in space and time}

\author{A. Kshetrimayum}

\affiliation{\mbox{Helmholtz Center Berlin, 14109 Berlin, Germany}}
\affiliation{Dahlem Center for Complex Quantum Systems, Freie Universit{\"a}t Berlin, 14195 Berlin, Germany}

\author{J. Eisert}

\affiliation{Dahlem Center for Complex Quantum Systems, Freie Universit{\"a}t Berlin, 14195 Berlin, Germany}
\affiliation{\mbox{Helmholtz Center Berlin, 14109 Berlin, Germany}}

\author{D. M. Kennes}
\affiliation{Institut f\"ur Theorie der Statistischen Physik, RWTH Aachen, 
52056 Aachen, Germany and JARA - Fundamentals of Future Information Technology}
\affiliation{Max Planck Institute for the Structure and Dynamics of Matter and Center for Free-Electron Laser Science, 22761 Hamburg, Germany}

\begin{abstract} 
The compelling original idea of a time crystal has referred to a structure that repeats in time as well as in space, an idea that has attracted significant interest recently. While obstructions to realize such structures became apparent early on, focus has shifted to seeing a symmetry breaking in time in periodically driven systems, a property of systems referred to as discrete time crystals. In this work, we introduce Stark time crystals based on a type of localization that is created in the absence of any spatial disorder. We argue that Stark time crystals constitute a phase of matter coming very close to the original idea and exhibit a symmetry breaking in space and time. Complementing a comprehensive discussion of
the physics of the problem, we move on to elaborating on possible 
practical applications and argue that the physical demands of witnessing genuine signatures of many-body localization 
in large systems may be lessened in such physical systems.
\end{abstract}

\date{\today} 
\maketitle

\section{Introduction}
Time crystals are structures that repeat periodically in both space and time \cite{SondhiReview,doi:10.1146/annurev-conmatphys-031119-050658}. While phases of matter that spontaneously break the continuous spatial translations symmetry in such a way are ubiquitous in nature, the corresponding breaking of time translation symmetry has been met with skepticism despite its earlier proposal by Wilczek in 2012~\cite{PhysRevLett.109.160401}. An early
setback to this idea came in the form of a 
no-go theorem by Oshikawa
~\cite{PhysRevLett.114.251603,2ndNoGo} which states that ground states of local Hamiltonians cannot host such symmetry breaking in time. Two loop-holes remain, however: abandoning the assumptions of a local interaction or that of being in equilibrium. Indeed, when giving up the former \cite{PhysRevLett.123.210602} a continuous time crystal can be found, albeit at the expense of highly intricate and engineered and largely unphysical long-ranged interactions. In the latter case, one resorts to driving the system periodically which explicitly breaks the continuous time-translation invariance down to a discrete one. However, this discrete symmetry can further be broken spontaneously, e.g., by period doubling \footnote{Period doubling is well known from classical non-linear mechanics, but finding it in a quantum system where systems evolve according to a linear unitary evolution is intriguing.} or beyond~\cite{PizziTC2019,PizziTC2019two}, and a so-called {\it discrete time-crystal} is found \cite{PhysRevLett.116.250401,PhysRevB.93.245145,PhysRevB.93.245146,PhysRevLett.116.250401,PhysRevLett.118.030401,PhysRevB.93.245145}, going back to the
seminal work Ref.\ \cite{PhysRevLett.116.250401}.
This second pathway of resorting to non-equilibrium 
has been realized experimentally \cite{1408.5148,ngupta_Silva_Vengalattore_2011,MonroeTimeCrystal,LukinTimeCrystal}  with a particular focus on one spatial dimension \cite{PhysRevA.91.033617,PhysRevLett.118.030401,Else16}, but with significant steps towards the
numerical exploration of such phases in two
spatial dimensions having been taken
\cite{Augustine2DTC}.

However, in all the realizations above, the focus has shifted to finding evidence of translation symmetry breaking {\it only in time}. And there are good reasons for this. The only practical way to overcome the no-go theorem as we discussed above is periodic driving. With periodic driving comes runaway heating and thermalization to infinite temperature in the generic case \cite{floquet_alessio}. Such an infinite temperature state cannot host interesting time crystalline phases.  A known way to overcome heating was put forward in the context of many-body localization \cite{huse2014phenomenology,RevModPhys.91.021001}, where localization even in the presence of interactions is induced by quenched, quasi-periodic or binary disorder \cite{MBL1,MBL2,Pollmann_unbounded,Prosen_localisation,1409.1252,Schreiber17,Luschen17,Bar_Lev_2017,Enss17,Kennes18,Augustine2DMBL}. With runaway heating suppressed in these situations discrete time-crystals can emerge. However, the use of such strong disorder, particularly for the purpose of time crystals is subtle. The disorder configurations are realization dependent due to which it has to be averaged over many number of draws. This increases the numerical and experimental effort and makes the setting less reliable.

Even more importantly, conceptually speaking, 
spontaneous symmetry breaking in space in such disordered lattices is not well defined and as a consequence the spatial symmetry breaking in discrete time crystals has not gathered much attention. Only when one does not
break spatial symmetry ``by hand'', one can hope to observe
genuine {\it symmetry breaking both in space and time}, conceptually
an intriguing situation. Another problem arising from disorder is the issue of the {\it instability} of such disordered configurations due to the presence of rare ergodic regions in higher dimensions as well as in systems with interactions that decay slower than exponential in space~\cite{Roeck2017,RoeckImbrie}. 

All of these points lead us to the pressing question of whether one can have a clean and more stable setting where the concept of {\it symmetry breaking both in space and time can be realized}, at least for long
times in a pre-thermal sense of the
term and possibly for all times.
In this work, we answer this question to the affirmative by introducing {\it Stark time crystals} -- discrete time crystals protected from runaway heating by Stark many-body localization, a mechanism we will briefly outline next. 

\section{Wannier-Stark localization}

{\it Wannier-Stark localization} -- the notion that non-interacting particles on a lattice will be localized upon applying a linear potential -- implies many confounding consequences \cite{PhysRev.117.432}. Recently, it was shown that this localization can survive even in the presence of interactions \cite{vanNieuwenburg9269,PhysRevLett.122.040606,PhysRevX.10.011042,Bhakuni_2020JOP,Bhakuni_2020PRB}. Therefore, such a system displays many similarities to a many-body localized one featuring genuine disorder.
Wannier-Stark-like localization in the presence of interactions was thus dubbed Stark many-body localization \cite{vanNieuwenburg9269,PhysRevLett.122.040606}. 
One of the similarities that is inherited by Stark many-body localized systems is barred    thermalization and
 we show here that by a similar suppression of runaway heating as in the many-body localized case a time crystalline structure can be induced.

Stark many-body localization  can thus be exploited to realize a quantum time crystal in the absence of any disorder and we explicitly construct such a {Stark quantum time crystal} and study its stability. First steps in this direction, albeit for a few spins and concentrating on cases where energy gradients and quenched disorder coexists have been undertaken in Ref.~\cite{PhysRevB.101.115303}. In contrast, here we aim at
exploring the spatio-temporal structure of the time-crystalline phase (for which a large number of spins is required) and aim to explore the dynamics in a purely Stark many-body localized time-crystal without any disorder terms. By this we add a localized time-crystal without spatial disorder or quasi-periodic potentials to the list of accessible phases of matter \cite{Russomanno17,Ho17,Rovny18,Augustine2DTC}. We report that the spatio-temporal behaviour of such a Stark time crystal is intriguingly rich and displays filaments of local period doubling extending further and further in time and finally condensing at the critical point where the time-crystalline behaviour is stabilized. Such spatio-temporal features have been inaccessible previously due to the requirement of disordered lattice potentials (outside of alternative routes to circumvent heating such as the presence of symmetry~\cite{HeylTC2020} or dissipation
~\cite{GambettaPRLTC2019,GambettaPRETC2019}). We furthermore identify an effect unique to Stark quantum time crystals: By resonance of the external drive  with the potential difference defining Stark localization, the drive which induces the time-crystal coherently destroys \cite{1912.09487} the localization and therefore the time-crystalline behaviour itself (at large enough potential difference this effect can be linked to the few-spin dynamics reported in Ref.\ \cite{PhysRevB.101.115303}). This leads to a periodic weakening of the time crystal state when the potential difference is tuned to integer multiples of the driving frequency. Furthermore, we discuss technological applications and how time crystalline behaviour can actually be used as a witness to distinguish a many-body localization from Anderson localization.

\begin{figure}[t!]
\centering
\includegraphics[width=.9\columnwidth]{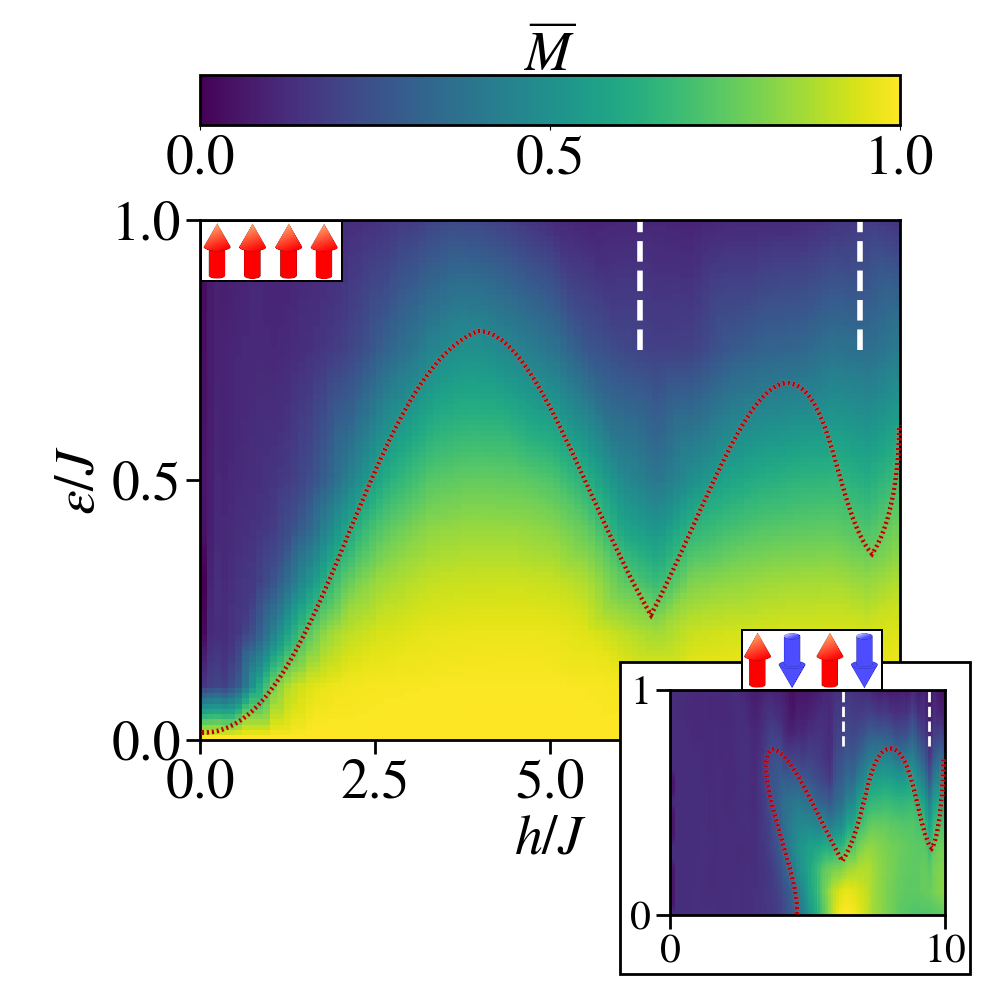}
\caption{Phase diagram of the discrete time-crystalline phases for the initial state being either ferro- (main panel) or anti-ferromagnetic (lower right inset). The false color plot shows $\overline M$, a quantity measuring the strength of the symmetry breaking in time (see Eq.~\eqref{eq:def_M} for the definition), in dependence of the potential difference $h$ and the imperfection of drive $\epsilon$. A dotted red line serves as a guide to the eye separating regions of rigid symmetry breaking from those where no rigid time-crystalline behaviour is found (quantified by $\overline M=0.5$). 
At $h/J$ being $2\pi$ or $3\pi$ (white vertical dashed line), the time-crystalline phase is weakened substantially. This is due to the coherent destruction of the localization by the external drive. The other parameters are $U/J=1$, $L=100$ and $JT/2=JT_1=JT_2=1$.}
\label{fig:Fig_PD}
\end{figure}

The main findings are summarized in the phase diagram of Fig.~\ref{fig:Fig_PD} for two different initial states, being either ferro- (main panel) or anti-ferromagnetic (inset). The quantity $\overline M$ shown in the false color plot, defined in Eq.~\eqref{eq:def_M}, quantifies the rigidity of the time crystal (see below).  We find that for sufficiently large potential difference $h$ the time-crystal phase can be stabilized by Stark many-body localization even for a non-zero perturbation in the driving $\epsilon$ (defined below). However, when the potential difference  coincides with an integer multiple of $\Omega=2\pi/T=\pi J$ highlighted by dashed white lines for $2\pi$ and $3\pi$, we find coherent self-destruction of the time-crystalline behaviour and $\overline M$ is decreased drastically.

\begin{figure*}[t!]
\centering
\includegraphics[width=.9\textwidth]{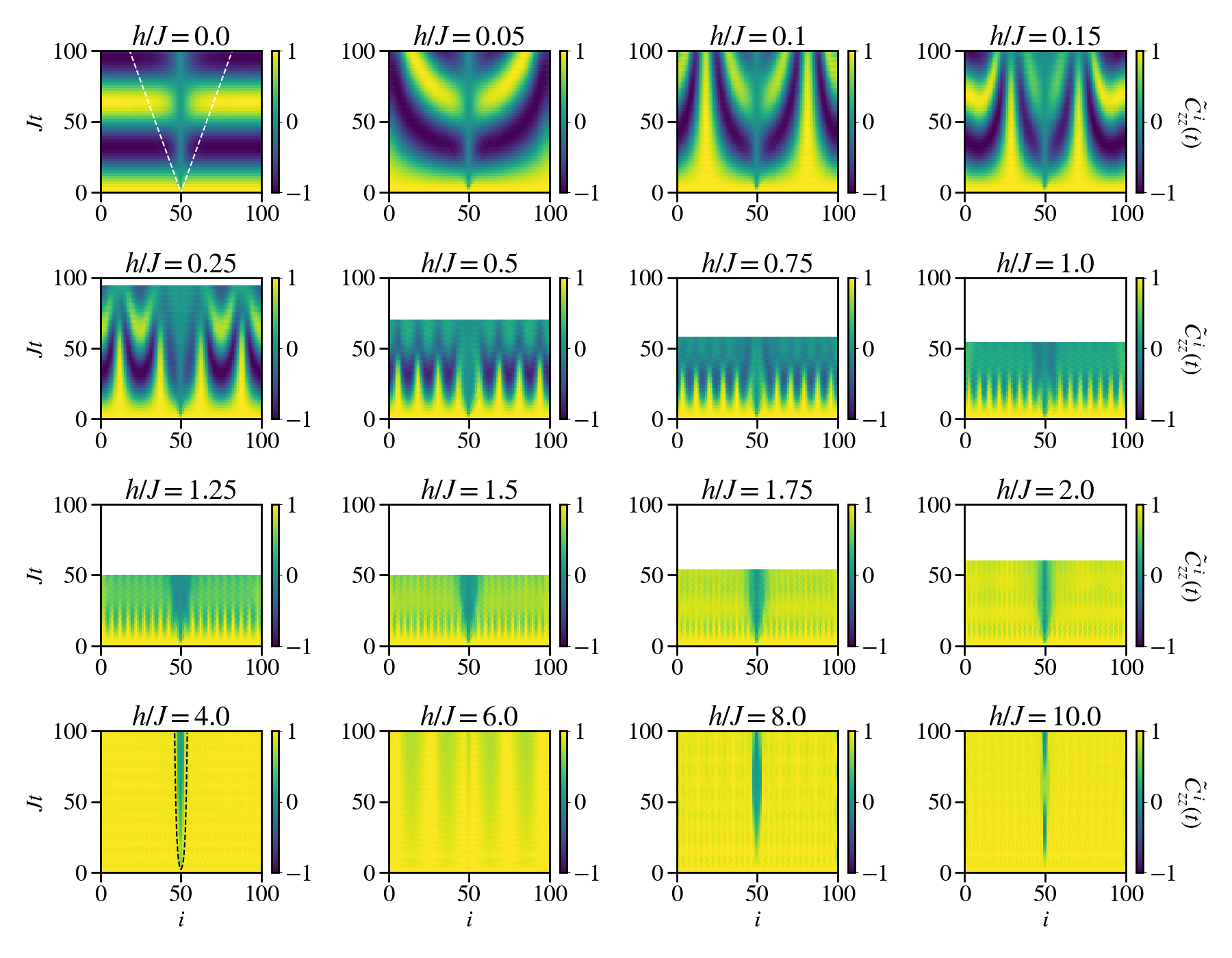}
\caption{Spatio-temperal structure of the symmetry breaking for a domain-wall initial state.  We consider the modified correlation function $\tilde C_{zz}(t)$ which in time remains close to one for a perfect time-crystal (yellow in false color plot). Deviations from yellow show that the time-crystal structure is broken. At $h/J=0$ no time-crystal forms due to the lack of Stark many-body localization. Ballistic jets  are found propagating from the domain-wall kink highlighted  by a white dashed line. Those jets are barely visible on a linear color scale, but very pronounced on a logarithmic one, see also the appendix. As the  energy-gradient $h$, localizing the system, is increased filaments of time-crystalline structure (yellow structures extended in time) are found with a filament density linearly increasing with $h$. When the density of filaments fills the entire system in space a time-crystal is established (compare also the appendix). At this point the transport originating from the domain wall is logarithmic (highlighted by dashed black line for $h/J=4$). For values of $h$ close to integer multiples of $\Omega=2\pi/T=\pi J$ (such as $h/J=6$), the time crystal is weakened again, which relates to coherent self-destruction. The other parameters are $U/J=1$, $L=100$, $\epsilon/J=0.2$ and $JT/2=JT_1=JT_2=1$.
}
\label{fig:Fig_phase_slip_ana}
\end{figure*}

\begin{figure}[t]
\centering
\includegraphics[width=.9\columnwidth]{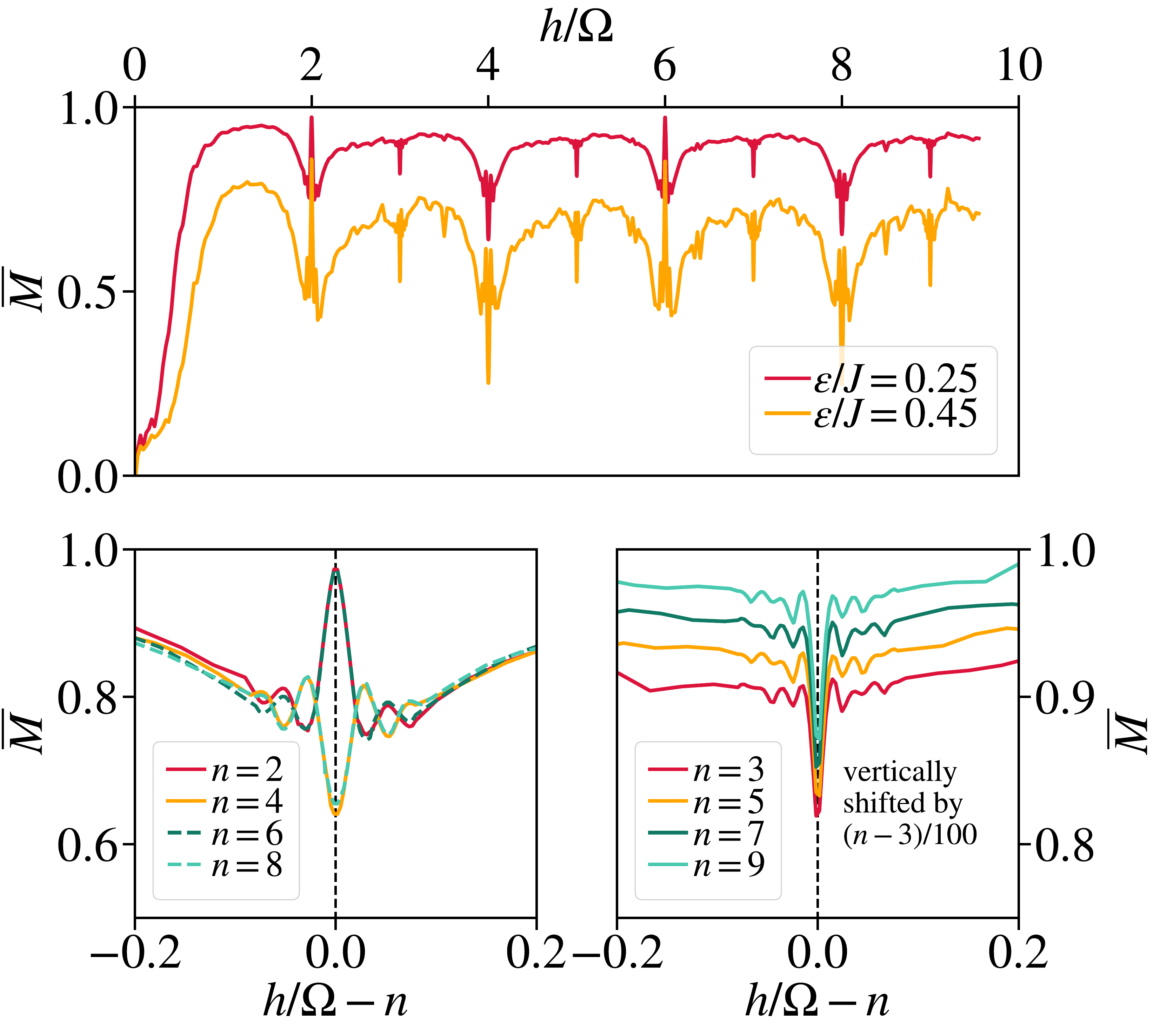}
\caption{Coherent self-destruction of a Stark time crystal. Top Panel: the time-averaged strength of the time-crystal $\overline M$ in dependence of the localization strength $h$ in units of the drive frequency $\Omega=2\pi/T$ for two different drive imperfections $\epsilon$. Bottom panels: Zoom into the region where $h$ is an even (left) or odd (right) multiple of $\Omega$. As $h$ approaches integer multiples of $\Omega=2\pi/T=\pi J$ the time-crystal structure is weakened in a resonant fashion.  The weakening is stronger for $h$ being an even than for $h$ being an odd integers of $\Omega$. The other parameters are $U/J=1$, $L=100$ and $JT/2=JT_1=JT_2=1$. 
}
\label{fig:Fig_coherent_destruction}
\end{figure}

\section{Model}
Following the work of Ref.~\cite{PhysRevLett.118.030401}, we 
aim at probing the existence of a Stark quantum time crystal in a one-dimensional chain of $L$ spin-$1/2$ particles by considering a stroboscopic Floquet Hamiltonian with period $T=T_1+T_2$. The time evolution operator 
for one period is taken as $U(T)=e^{-i H_{\rm Flip}T_2}\cdot e^{-i H_{\rm Stark} T_1}$, with
\begin{equation}
H_{\rm Stark}=\sum
\limits_{i=1}^{L-1} 
\left(J S^x_{i}S^x_{i+1}+JS^y_{i}S^y_{i+1}+U S^z_{i}S^z_{i+1}\right)
    +h \sum\limits_{i=1}^L i S^z_{i}
\end{equation}
and 
$H_{\rm Flip}=\left(\pi/T_2-\epsilon\right) \sum_{i=1}^LS^x_i$.
This entails that the total Hamiltonian is time periodic with period $T$, where within one period of $t\in [0,T)$, first only $H_{\rm Stark}$ is active for times $t\in [0,T_1)$, while for the second part of the period $t\in [T_1,T)$ only $H_{\rm Flip}$ acts on the systems.
Here, $H_{\rm Stark}$ is the Hamiltonian that describes a nearest-neighbor interacting spin chain subject to a linear potential difference $h$ from site to site (non-linearities in this gradient are analyzed in the appendix). This part exhibits Stark many-body localization in the right parameter regime \cite{vanNieuwenburg9269,PhysRevLett.122.040606}. {It is also translationally invariant in the gauge choice where the hoppings are time dependent (adding another Floquet type driving to the system). But even without this gauge 
choice, the physics of the Hamiltonian is translation invariant in the sense that a particle is governed by the same local Hamiltonian when placed at site $i$ or site $i+r$ as the overall energy does not enter.} $H_{\rm Flip}$ describes a spin rotation around the x-axis. If $\epsilon=0$, 
this part of the Hamiltonian performs a perfect 180$^\circ$ rotation (flip) of the spin over the time $T_2$. A quantum time crystal is found if the breaking of the underlying discrete time-translation symmetry (e.g., period doubling) is {\it robust} to perturbations $\epsilon\neq 0$ (i.e., the deviation from the perfect flip). As initial state vectors, 
we test different configurations being either ferromagnetic $\left|\dots,\uparrow,\uparrow,\uparrow,\uparrow,\dots\right\rangle$, a domain wall $\left|\dots,\downarrow,\downarrow,\uparrow,\uparrow,\dots\right\rangle$ or anti-ferromagnetic $\left|\dots,\uparrow,\downarrow,\uparrow,\downarrow,\dots\right\rangle$. Starting from a ferromagnetic configuration we thus first introduce one phase slip (domain wall) and then an extensive number of $L/2$ phase slips (anti-ferromagnetic).

\section{Methods}
We solve the dynamics of the system starting from these different initial states using a numerically exact {\it density matrix renormalization group} approach set up in 
{\it matrix product states} 
\cite{MPSRev,Fannes-CMP-1992,MPSReps}. 
We propagate the system in real time using a fourth order Suzuki-Trotter decomposition with $J\Delta t =0.05$ and adaptive bond-dimension such that the summed truncated error throughout the total time-evolution remains below $10^{-5}$ for which converged results, on the scale of every plot, are obtained. For detecting the time crystalline behaviour, we broaden the notion of  {\it long-ranged order} of local order parameters $O(i,t)$, which for spatial crystals is present in {\it equal-time correlations in space} 
$C^t_{OO}(i,i')=\lim_{|L|\rightarrow\infty}\langle O(i,t) O(i',t)\rangle \neq 0$ 
to {\it equal-space correlations in time}  
\begin{equation}
C^i_{OO}(t,t')=\lim_{|L|\rightarrow \infty} \langle O(i,t) O(i,t') \rangle = f(t,t').
\end{equation}
 
Our system is called a time-crystal if (for $\epsilon\neq 0$) $f(t,t')$ shows a long time $t\gg t'$ non-trivial (ordered) behaviour that breaks the discrete time-translation symmetry of the drive of the Hamiltonian.
{We would like to point out here that our definition of the ordered parameter $f(t,t')$ is equivalent to the one introduced in Ref.\  \cite{PhysRevLett.114.251603} in the presence of long-ranged order for a Floquet time crystal.}
We then concentrate on the particular case where $O=S^z_i$ and $t'=0$.
The corresponding correlation function $C^i_{S^zS^z}(t,0)$ can now be written as $C^i_{zz}(t)$. For simplicity of depiction, we also introduce $\tilde C^i_{zz}(nT)=(-1)^n C^i_{zz}(nT)$ evaluated at stroboscopic times $t=nT$ ($n$ being an integer) as well as the space averaged quantities 
$ C_{zz}(t)=(1/L)\sum_i C
^i_{zz}(t)$ 
and analogously for $\tilde C_{zz}(t)$.
 The former definition of $\tilde C^i_{zz}(nT)$ is useful, as it monitors period doubling, staying close to $1$ when the time-crystal persists. We also introduce the time-averaged quantity (over $50$ periods) \footnote{The appropriate choice of averaged cycles depends of course on the parameters used. } 
\begin{equation}
\overline M  =\frac{1}{50}\sum_{n=0}^{50} \tilde C_{zz}(nT), \label{eq:def_M}
\end{equation}
which is a useful {\it single} number measure for the quality of the period-doubling time-crystal (being close to $1$ if a rigid time-crystal is found). 

\section{Results and discussion}  

The main result of the phase diagram of the time crystal, using $\overline M$ as a metric, is 
summarized in the discussion of Fig.~\ref{fig:Fig_PD} above. The  time-crystalline phase is a many-body effect and is absent without interactions (see also the appendix). In fact, this is one of the crucial features of our study. As we have seen, the many-body interaction is essential to see robust features of the time crystal. This means that Anderson localization alone is not sufficient 
to realize a time crystal. This leads us to believe that such 
robust features of time crystal can indeed be used as witness 
to distinguish  many-body localization from the Anderson 
localization case, just as it bears witness to the Wannier-Stark versus a Stark many-body localized case. As no programming of
disorder is required, such witnessing of many-body localization
may be significantly more feasible than one based on 
{\it logarithmic entanglement growth} 
\cite{Prosen_localisation,Pollmann_unbounded,GreinerMBL,ChiaroMBL}
or the {\it behaviour of two-point correlation functions} 
\cite{Accessible}. 
We complement these findings by considering the correlation functions $\tilde C_{zz}$ 
and $C_{zz}$  at stroboscopic and non-stroboscopic times more closely in the appendix.

Next, we reveal the intricate spatio-temporal structure of a Stark time-crystal, a major feature that was not explored before, due to the limited system size \cite{PhysRevB.101.115303} (for an analysis of system size effects, see the appendix). In Fig.~\ref{fig:Fig_phase_slip_ana}, we summarize results for the space resolved correlation function $\tilde  C_{zz}(nT)$ varying $h/J$ in a series of plots (for more data varying $\epsilon$ and $L$ see the appendix). We concentrate on the domain-wall initial state where we can clearly separate the features arising from the phase-slips in  the initial state and the dynamics of the (imperfect) spin-flips. In the case of $h=0$ ballistic jets emanate from the phase slip (seen clearly only on a log scale, see the appendix), while with the onset of Stark many-body localization this transport becomes blocked. This behaviour is superimposed by the way the time-crystal emerges at increased $h$: filaments of time-crystalline structure (yellow structures extended in time) extend over longer times and their density increases  linearly with $h$. As the density of filaments condenses in  space, a time-crystal with also spatial symmetry breaking is established (see also the appendix for further analysis). However, we also report here that for $h$ close to integer multiples of $\Omega=2\pi/T$ (such as $h/J=6$), the time crystal is weakened again and darker regions ($\overline M$ deviates from 1) show up in the time crystal.

This effect relates to coherent self-destruction and is summarized in  Fig.~\ref{fig:Fig_coherent_destruction}. We show horizontal slices of Fig.\ref{fig:Fig_PD} for two value of $\epsilon$ and a domain-wall initial state. Here, we scale $h$ with respect to $\Omega$, which highlights the fact that at integer multiples of $h=n\Omega$, the time-crystal is weakened. We can understand the reported  behaviour by using Floquet theory in which copies of the original Hamiltonian but shifted in energy by integer multiples of $\Omega$ need to be considered \cite{KUWAHARA201696,Eissing16}. In this language, levels that are $n$ sites apart from each other (with energy difference of $nh$) become resonant in the $n$-th Floquet replica. Another way to view this is to consider, that a particle at one site can absorb $n$ ``photons'' of the driving field and is then resonant with a level being $n$ sites away. Such a resonant process destroys localization and thus the time-crystalline behaviour \cite{1912.09487,PhysRevB.101.115303}. A more detailed study of the even-odd effects reported in the coherent self-destruction shown in Fig.~\ref{fig:Fig_coherent_destruction} is left for future work.

We would like to point out that the above discussed spatio-temporal features are absent in the case of a time crystal realized with disorder (shown in the appendix). To the best of our knowledge such a symmetry breaking in both space and time has not been realized before rendering our setting closer to the original proposal of time crystals. In the future, it would be analyze to associate an order parameter that is defined in both space and time.

\section{Conclusion and outlook} 
We have analyzed the emergence and stability of a Stark quantum time-crystal  where the many-body localization is induced solely by a linear potential. Since our clean setting does not involve any disorder, we can find symmetry breaking emerging in both space and time. We find that Stark quantum time crystals showing spontaneous period doubling can be robust as long as the potential difference from site to site remains off resonant with processes arising from absorptions of the underlying drive. We have also discussed the possibilities of using time crystals as MBL witness. We will explore this in more detail in the future.
Another interesting route of future study should address the question whether using such a Stark quantum time crystal, novel quantum pumps (attaching reservoirs to the time crystal structure) can be realized and whether such pumps are, e.g., useful in 
{\it quantum metrological applications}: It has been
suggested that a time crystal could beat the Heisenberg limit
in quantum metrology \cite{Metrology2}, an application significantly
more plausible in the absence of disorder. 
It is the hope that the present work not only
sheds new light into the conceptual foundations
of symmetry breaking in space and time, but also 
stimulates such technological applications.

\section{Acknowledgments} 
Funded by the Deutsche Forschungsgemeinschaft (DFG, German Research Foundation) under Germany's Excellence Strategy - Cluster of Excellence Matter and Light for Quantum Computing (ML4Q) EXC 2004/1 - 390534769
under CRC 183 (Project B01) and EI 519/15-1. It has also been 
supported by the Templeton Foundation and has 
received funding from the European Unions Horizon 2020 research and innovation programme under grant agreement No 817482 (PASQuanS).
We acknowledge support from the Max Planck-New York City Center for Non-Equilibrium Quantum Phenomena.

\section{Appendix}

\begin{figure}[h!]
\centering
\hspace{0.3cm}\includegraphics[width=0.225\columnwidth]{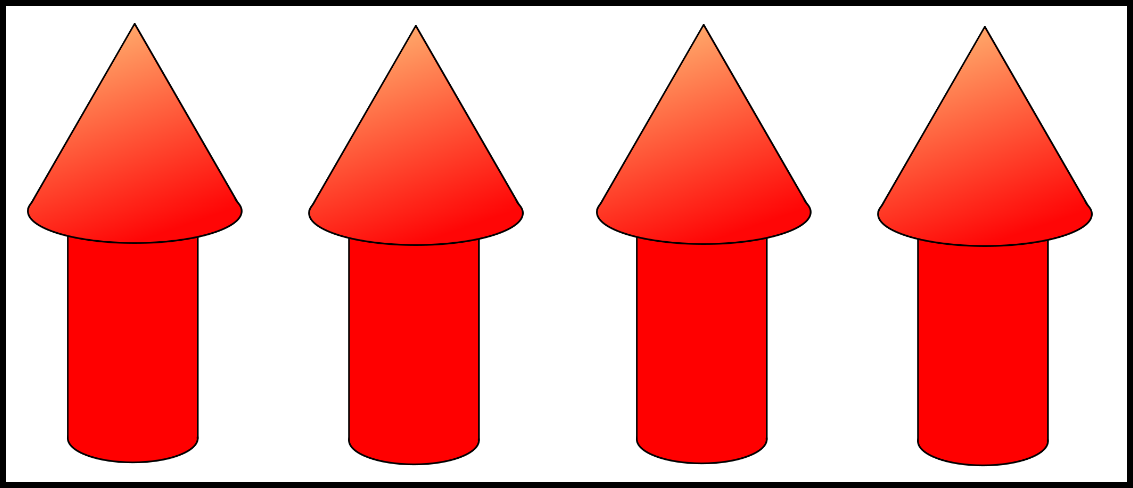}
\hspace{0.75cm}\includegraphics[width=0.225\columnwidth]{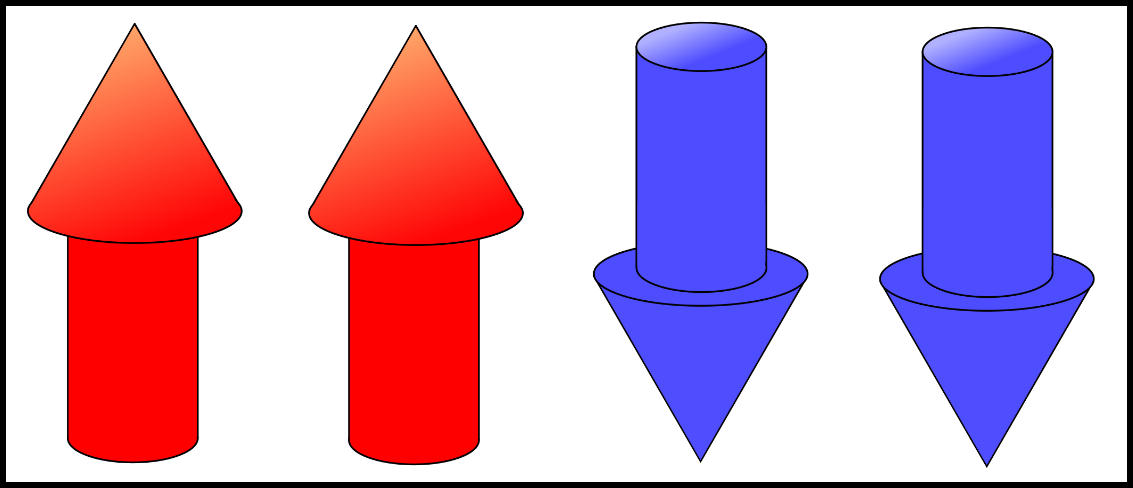}
\hspace{0.65cm}\includegraphics[width=0.225\columnwidth]{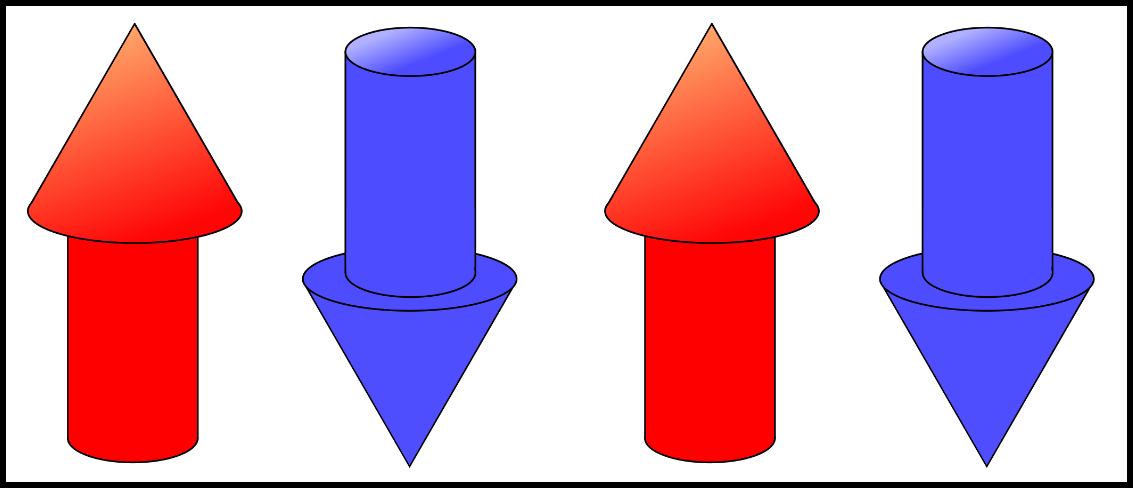}

\includegraphics[width=.6\columnwidth]{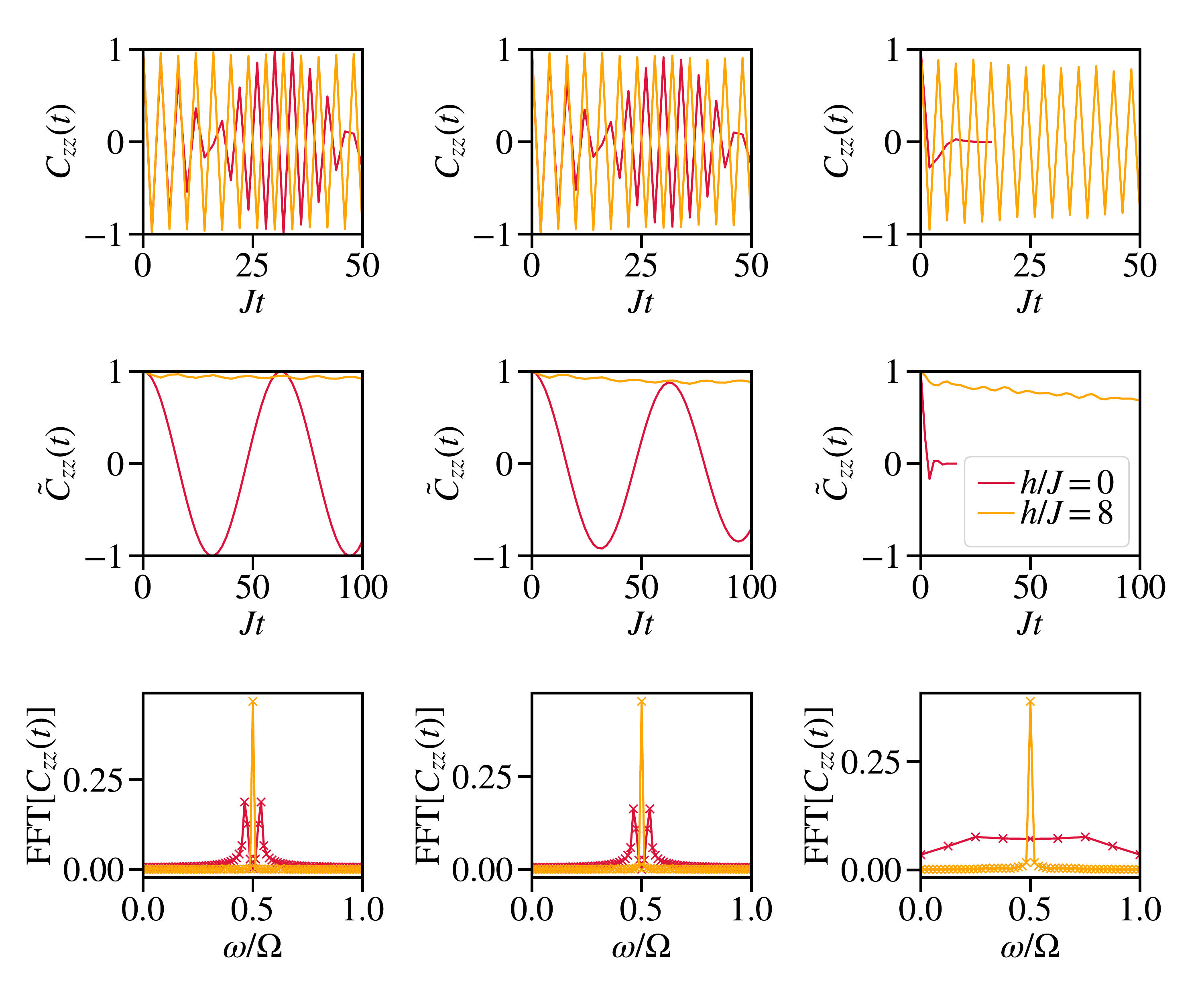}
\caption{Comparison of the time-crystalline behaviour for a ferromagnetic, domain-wall or anti-ferromagnetic initial state (left to right columns) for two values of $h/J=0$ and $h/J=8$. The first row shows the different initial states considered. The second row gives the correlation function $C_{zz}(t)$ which displays the spontaneous period doubling only when the system is Stark  many-body localized ($h/J=8$). The third row demonstrates the modified  correlation function $\tilde C_{zz}(t)$ measuring the stability of the period doubling (time-crystal). The fourth row displays the fast Fourier transform of $C_{zz}(t)$. Period doubling (signal at frequencies $\omega=\Omega/2$) is robust for all initial states only in the presence of Stark many-body localization ($h/J=8$).  The other parameters are $U/J=1$, $L=100$ and $JT/2=JT_1=JT_2=1$. 
}
\label{fig:Fig_C}
\end{figure}

This appendix provides additional data on Stark quantum time crystals. 
In Fig.~\ref{fig:Fig_C}, we compare the dynamics found for the three different initial states considered (three different columns of the figure). In the second and third row we show $ C_{zz}(nT)$ and $\tilde C_{zz}(nT)$, respectively. These rows indicate that for $h=0$ a (decaying) beating pattern in $ C_{zz}(nT)$ and a oscillating/decaying form for  $\tilde C_{zz}(nT)$ is found. In marked contrast for $h/J=8$ a time-crystalline behaviour is stabilized by Stark many-body localization (for more data on different interactions strength see the appendix). $ C_{zz}(nT)$ shows a robust period doubling and $\tilde  C_{zz}(nT)$ stays close to $1$. The fourth row shows the fast Fourier transform of $ C_{zz}(nT)$. In the case of a time crystal ($h/J=8$) we find a robust frequency feature at $\Omega/2$ (period doubling), while at $h/J=0$ this peak is not robust to perturbations and splits up.

\begin{figure}[h!]
\centering
\includegraphics[width=.6\columnwidth]{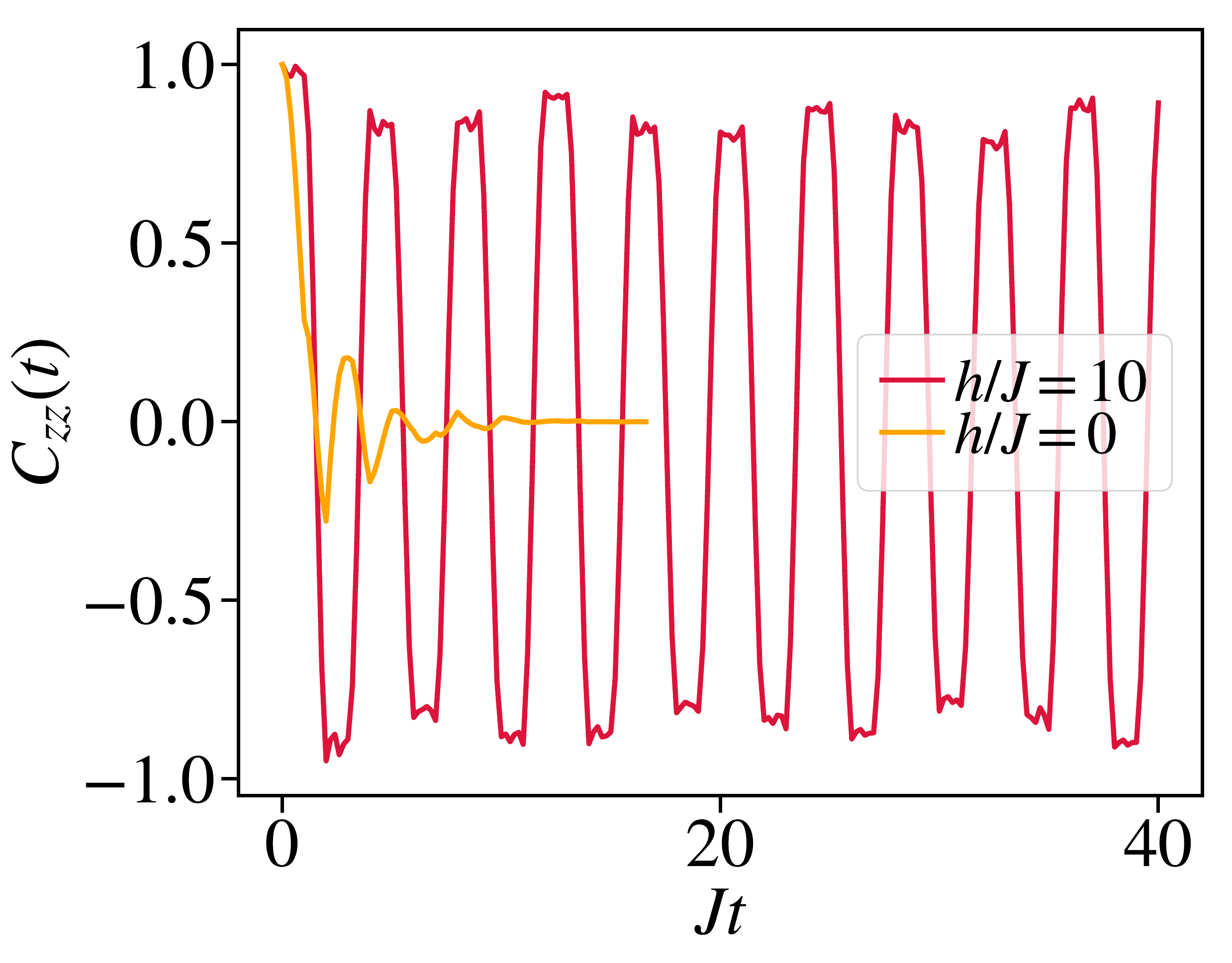}
\caption{Typical time evolution showing all times calculated (instead of only the stroboscopic ones which we concentrate on in the main text) of the correlation function $ C_{zz}(t)$  for two values of $h/J$ being in the time-crystalline or trivial region.  The other parameters are $U/J=1$, $L=100$, $\epsilon/J=0.2$ and $JT/2=JT_1=JT_2=1$. 
}
\label{fig:Supp_Fig_fullt}
\end{figure}

\begin{figure}[h!]
\centering
\includegraphics[width=.6\columnwidth]{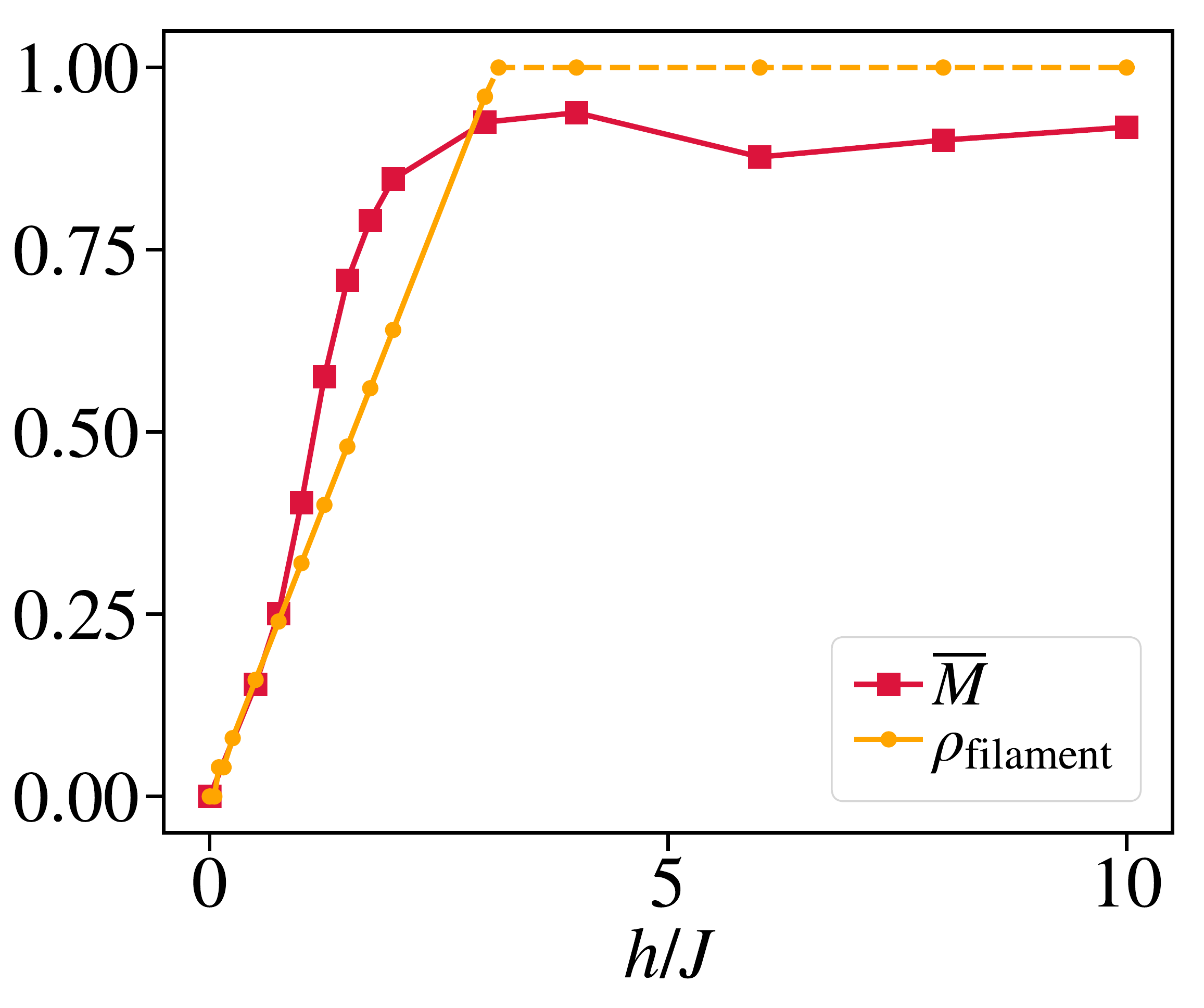}
\caption{Comparison between the filament density $\rho_{\rm filament}$, defined by the number of filaments found in Fig.~2 of the main text divided by half the system size, and the quantity $\overline{M}$ quantifying the stability of the time crystal. As the density of filaments $\rho_{\rm filament}$ approaches unity the time-crystal condenses and $\overline{M}\approx 1$. The parameters are the one of Fig.~2 of the main text.
}
\label{fig:Supp_Fig_PL}
\end{figure}

\begin{figure}[!htb]
\centering
\includegraphics[width=.8\columnwidth]{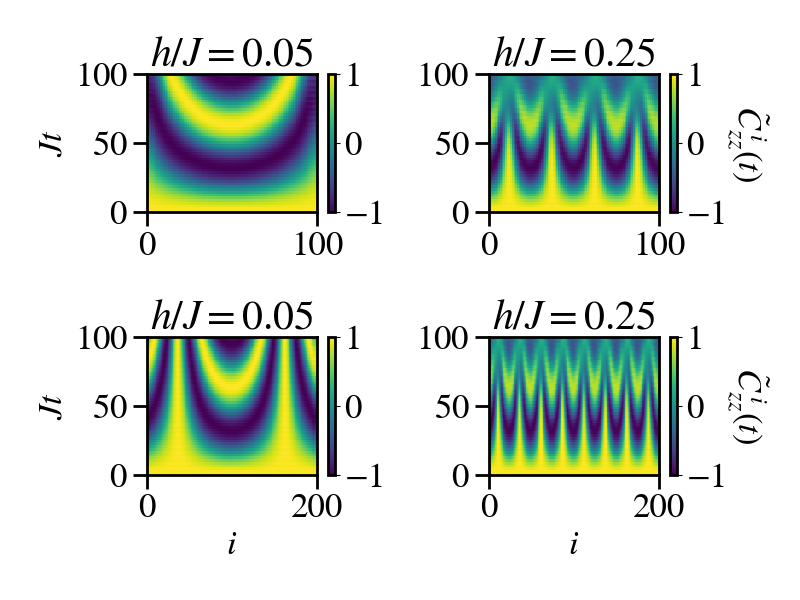}
\caption{The same as in Fig.~2 of the main text, but using from a ferromagnetic initial state instead of a domain wall and comparing two system sizes $L=100$ (top row) and $L=200$ bottom row, for two values of $h/J=0.05$ (left column) and $h/J=0.25$ (right column).
}
\label{fig:Supp_Fig_phase_slip_ana_L}
\end{figure}

\begin{figure}[h!]
\centering
\includegraphics[width=.6\columnwidth]{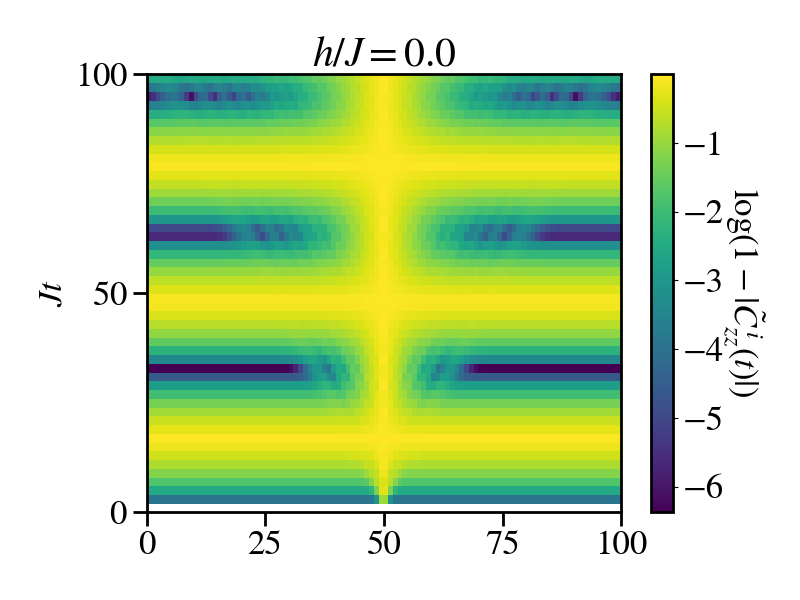}
\caption{The same as in the topmost left panel of Fig.~2 of the main text, but instead of showing $\tilde C^i_{zz}(t)$ we show the log of the absolute value of one minus this quantity. In this way of depiction of the data, the ballistic jets mentioned in the main text are clearly visible.  
}
\label{fig:Supp_Fig_phase_slip_ana_lightcone}
\end{figure}

\begin{figure}[h!]
\centering
\includegraphics[width=.6\columnwidth]{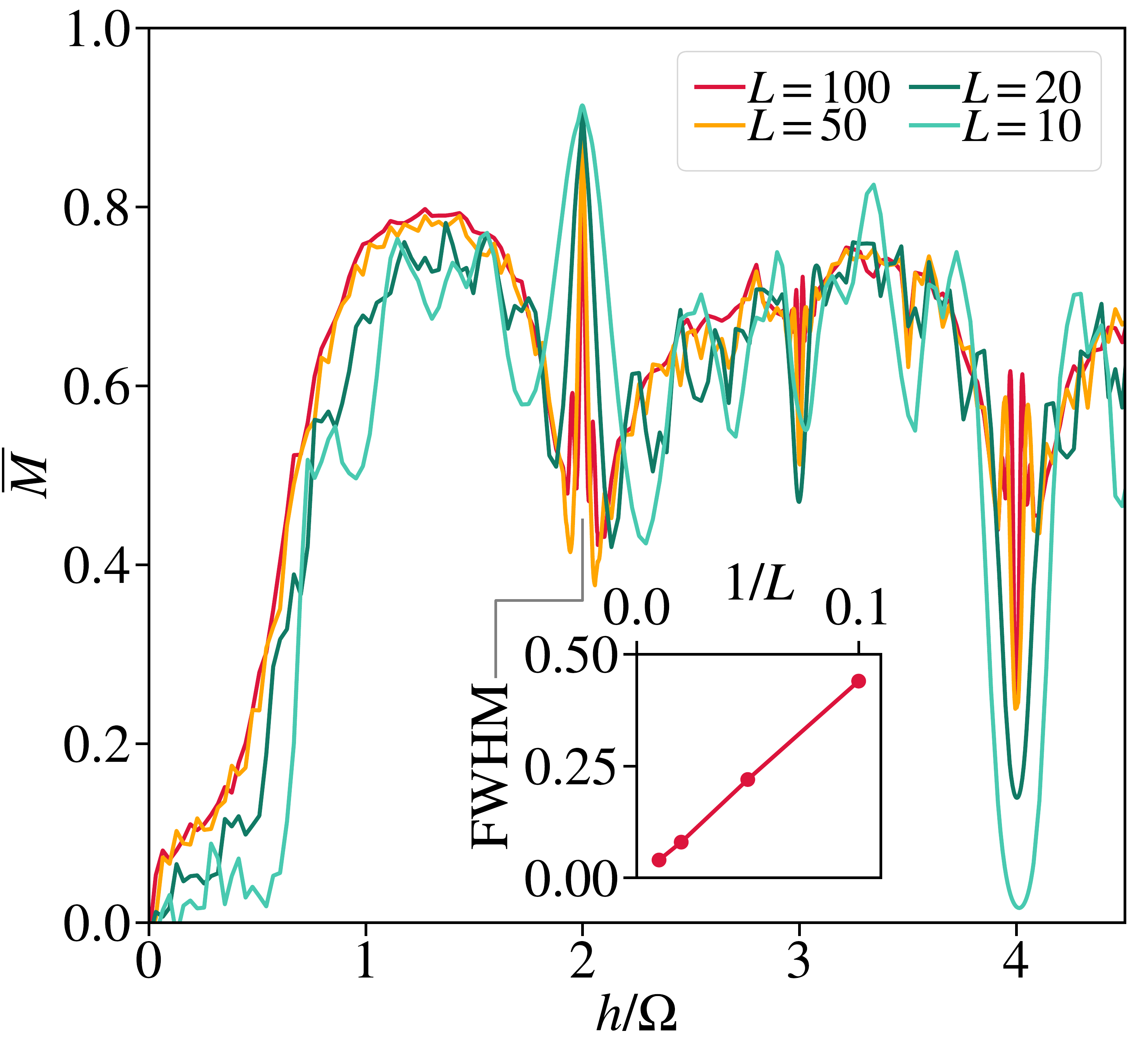}
\caption{Main panel: Same as Fig.~3 of the main text but for different system sizes $L$. Inset: Full width at half maximum (FWHM) of the peak centered at $h/\Omega=2$. System sizes of $L\approx50$ are needed to correctly capture the transition at small $h/\Omega$. The central peak structures relating to coherent self-destruction scale as $1/L$. At small system sizes $L\approx10$ the central features are washed out to large degree.  
}
\label{fig:Supp_Fig_coherent_des_L.pdf}
\end{figure}

\begin{figure}[h!]
\centering
\includegraphics[width=.6\columnwidth]{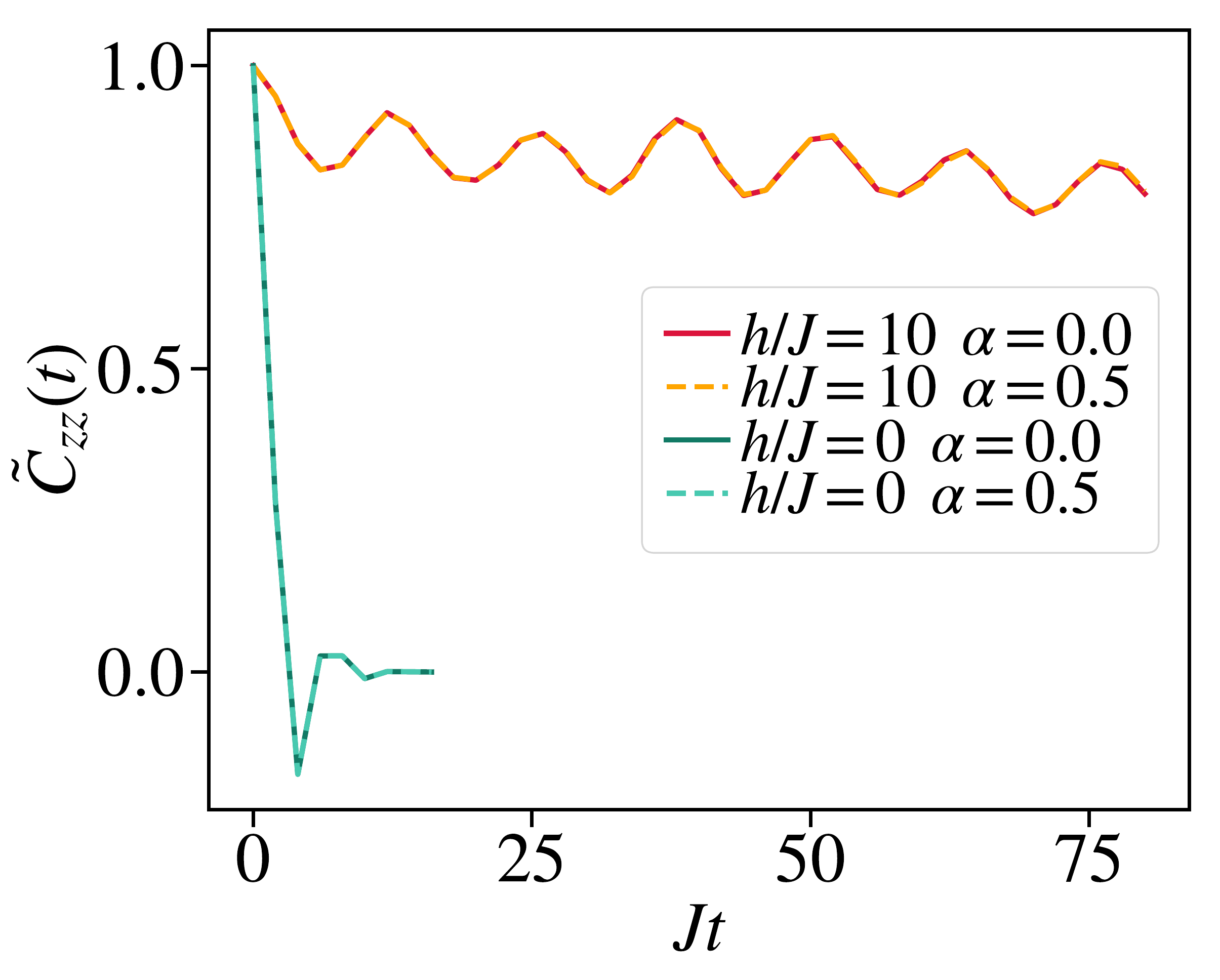}
\caption{Stability of the results with respect to variations in the linearity of the gradient $\alpha$. The findings are insensitive.  The other parameters are $U/J=1$, $L=100$, $\epsilon/J=0.2$ and $JT/2=JT_1=JT_2=1$.
}
\label{fig:Supp_Fig_alphavar}
\end{figure}

\begin{figure}[h!]
\centering
\includegraphics[width=.6\columnwidth]{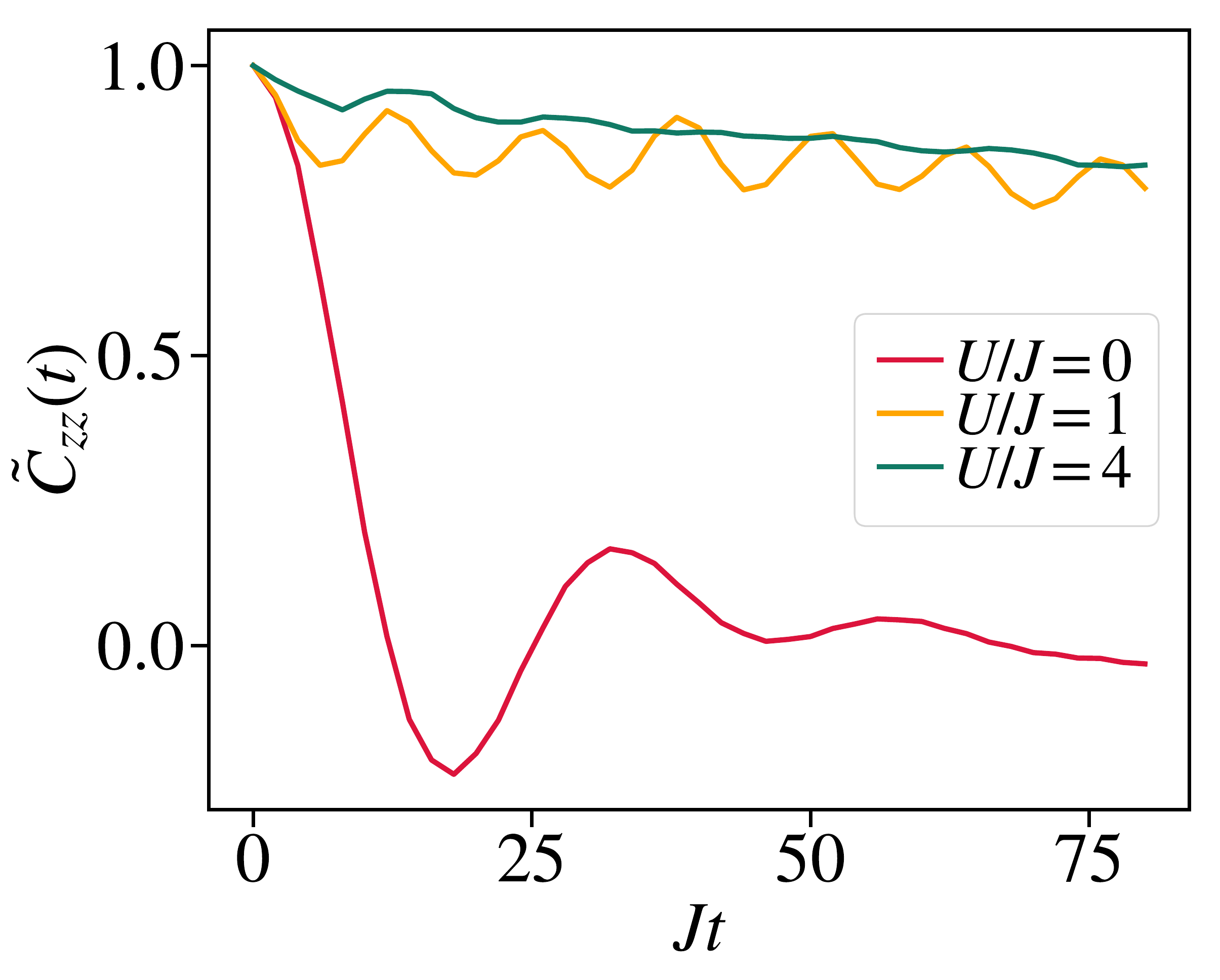}
\caption{Demonstration that the time-crystalline behaviour is interaction induced.  The other parameters are $h/J=10$, $L=100$, $\epsilon/J=0.2$ and $JT/2=JT_1=JT_2=1$.
}
\label{fig:Supp_Fig_Uvar}
\end{figure}

In Fig.~\ref{fig:Supp_Fig_fullt} we show the correlation function $C_{zz}(t)$ studied in the main text only at stroboscopic times  also at non-stroboscopic times. In Fig.~\ref{fig:Supp_Fig_alphavar} we numerically substantiate the claim, made in the main text, that small non-linearities in the energy gradient do no affect the time-crystalline behaviour reported. For this we add the term
\begin{equation}
    H_{\rm non-lin}=\alpha J \sum\limits_{i=1}^L (i/L)^2 S^z_{i}
\end{equation}
to $ H_{\rm Stark}$ and compare $\alpha=0$ to $\alpha\neq 0$. In Fig.~\ref{fig:Supp_Fig_Uvar} we study the influence of varying $U$. For $U=0$ where the system can be mapped to non-interaction fermions by a Jordan-Wigner transformation \cite{Jordan1928}, the time-crystalline behaviour is lost.

\begin{figure}[h!]
\includegraphics[width=\columnwidth]{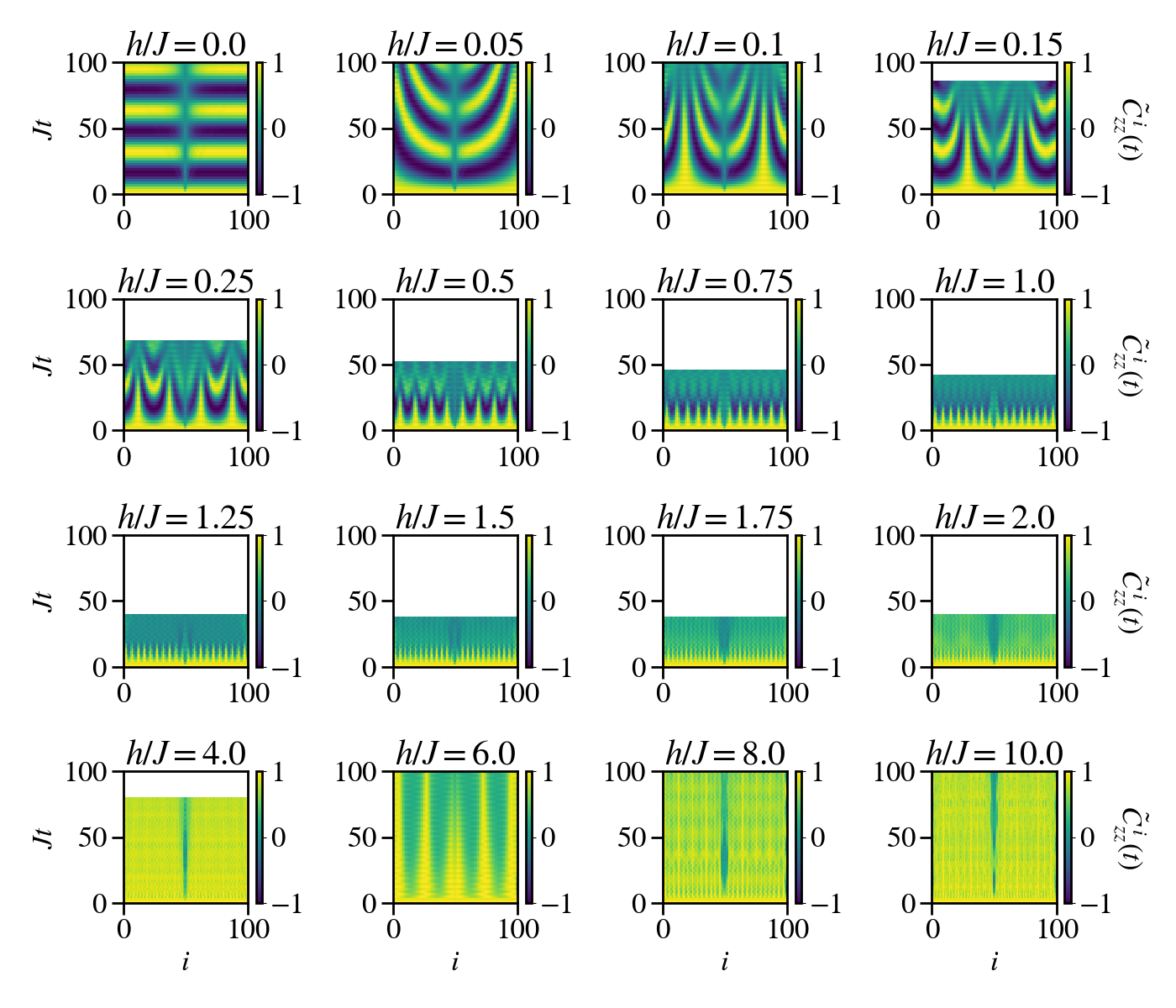}
\caption{Same as Fig.~2 of the main text but with $\epsilon/J=0.4$.
}
\label{fig:Supp_Fig_phase_slip_ana_eps}
\end{figure}

Fig.~\ref{fig:Supp_Fig_phase_slip_ana_eps} provides the same as  Fig.~2 of the main text but for larger $\epsilon/J=0.4$. Fig.~\ref{fig:Supp_Fig_PL} shows the density of filaments found in Fig.~2 of the main text compared to $\overline M$ showing that when $\overline M$ approaches unity, signalling the onset of the time crystal, the density of filaments approaches unity as well. Before that the density of filaments increases linearly with $h$. Fig.~\ref{fig:Supp_Fig_phase_slip_ana_L} shows a comparison of two different values of the system size $L$ and Fig.~\ref{fig:Supp_Fig_phase_slip_ana_lightcone} shows the light cone, ballistic jets reported for $h=0$ in Fig.~2 of the main text on a logarithmic scale.

\begin{figure}[h!]
\includegraphics[width=\columnwidth]{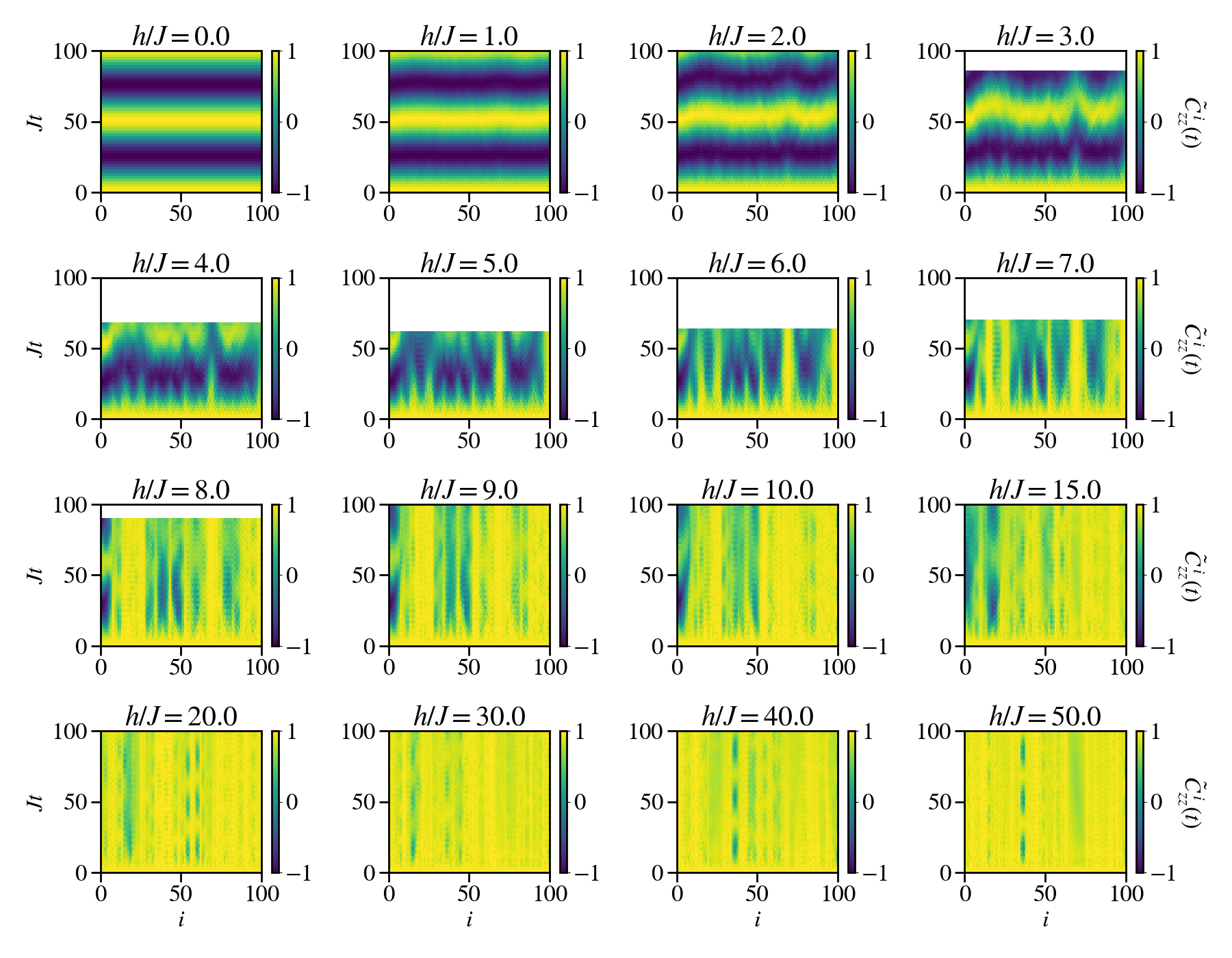}
\caption{Same as Fig.~2 of the main text but with $\epsilon/J=0.25$ and quenched disorder drawn uniformly from $[0,h)$.
}
\label{fig:Supp_Fig_phase_slip_ana_eps_MBL}
\end{figure}

Finally, Fig.~\ref{fig:Supp_Fig_coherent_des_L.pdf} shows the same as Fig.~3 of the main text but for different system sizes $L$. A system size of $L=50$ is needed to capture the transition point into the time crystalline phase correctly at small $h/\Omega$ and to clearly see the effects of coherent self-destruction. For comparison Fig.~\ref{fig:Supp_Fig_phase_slip_ana_eps_MBL} shows the same as Fig.~\ref{fig:Supp_Fig_phase_slip_ana_eps} but for quenched random disorder drawn uniformly from $[0,h)$.

%

\end{document}